\documentclass[preprint]{aastex}
 
\newcommand{\msun}{{\rm\,M_\odot}}
 
\begin{document}
 
\title{The Kozai Mechanism and the Evolution of Binary Supermassive
Black Holes}
 
\author{Omer Blaes, Man Hoi Lee, and Aristotle Socrates}
\affil{Department of Physics, University of California, Santa Barbara,
CA 93106}

\begin{abstract}
We consider the dynamical evolution of bound, hierarchical triples of
supermassive black holes that might be formed in the nuclei of galaxies
undergoing sequential mergers.  The tidal force of the outer black hole
on the inner binary produces eccentricity oscillations through the Kozai
mechanism, and this can substantially reduce the gravitational wave
merger time of the inner binary.  We numerically calculate the merger time
for a wide range of initial conditions and black hole mass ratios, including
the effects of octupole interactions in the triple as well as general
relativistic periastron precession in the inner binary.  The semimajor axes
and the mutual inclination of the inner and outer binaries are the most
important factors affecting the merger time.  We find that for a random
distribution of inclination angles and approximately equal mass black holes,
it is possible to reduce the merger time of a near circular inner binary 
by more than a factor of ten in over fifty percent of all cases.  We
estimate that a typical exterior quadrupole moment from surrounding matter
in the galaxy may also be sufficient to excite eccentricity oscillations
in supermassive black hole binaries, and also accelerate black hole mergers.
\end{abstract}
 
\section{INTRODUCTION}

The ubiquity of supermassive black holes (SMBHs) in the nuclei of many
galaxies (e.g. Magorrian et al. 1998) suggests that binary and multiple
SMBH systems may also be widespread.  For example, SMBH binaries should
form in galaxy mergers (Begelman, Blandford, \& Rees 1980), which is a
common process
in hierarchical models for galaxy formation and evolution (e.g. Kauffmann
\& Haehnelt 2000; Menou, Haiman, \& Narayanan 2001).  SMBH binaries have
been invoked to explain periodic
wiggles in extragalactic radio jets (e.g. Roos, Kaastra, \& Hummel 1993),
periodic flares in the BL Lac source OJ~287 (e.g. Lehto \& Valtonen 1996),
variations
in the apparent superluminal transverse velocities and position angles of the
3C~273 radio jet (Romero et al. 2000), and
the ``core-type'' nuclear surface brightness profiles of bright elliptical
galaxies (Ebisuzaki, Makino, \& Okumura 1991; Quinlan \& Hernquist 1997;
Faber et al. 1997; Milosavljevi\'c \& Merritt 2001).

SMBH binary mergers would be powerful sources of gravitational
waves (Thorne \& Braginsky 1976) which should be easily detectable from
planned space-based gravitational wave observatories such as the
{\it Laser Interferometer Space Antenna} (Bender et al. 1998).
However, it is still uncertain whether binary SMBHs can evolve to small
enough semimajor axes that gravitational radiation drives them to merge.
Energy can be extracted from a wide binary through interactions
with surrounding matter.  For example, encounters with passing stars will
shrink the binary, but this process will eventually be limited by the
depletion of stars on sufficiently radial orbits to encounter the ever
tightening binary.  Detailed work on this aspect of the problem has been
done by numerous authors (e.g. Quinlan 1996, Quinlan \& Hernquist 1997,
Milosavljevi\'c \& Merritt 2001).  A key uncertainty is the amount of
stochastic wandering that the binary undergoes at the bottom of the potential
well of the galaxy, allowing it to interact with many more stars.
Another important process that has received relatively less attention
is interaction with surrounding gas, which can also extract energy
from the binary and drive it to merge (e.g. Armitage \& Natarajan 2002
and references therein).

Large galaxies are typically the product of multiple mergers over the history
of the universe.  If the characteristic merger time of binary SMBHs
is not much less than the characteristic time scale between mergers, then
interactions with a third SMBH or a second SMBH binary will likely take
place.  Strong encounters within these multiple black hole systems may
lead to slingshot ejections of SMBHs from the nucleus of the galaxy
(Valtonen et al. 1994, Valtonen 1996).  On the other hand, the very existence
of the additional galaxy merger may introduce more stellar encounters with
the original central binary and drive it to merge  (Roos 1988).  Moreover,
repeated encounters between a third black hole and the inner binary can
increase the probability that the inner binary attains a large
eccentricity, thereby
accelerating the energy and angular momentum loss by gravitational radiation
(Makino \& Ebisuzaki 1994).

A modification of the last scenario is a case where the
third black hole has evolved to the point that it has become bound to the SMBH
binary, but has not yet come close enough for an unstable three-body
interaction
to take place.  The system then forms a hierarchical triple and can be treated
as consisting of an inner binary and an outer binary.  In that case,
if the mutual inclination angle between the inner and outer binaries
is high enough, then the time-averaged tidal gravitational force on the
inner binary can induce an oscillation in its eccentricity.
This effect is known as the Kozai mechanism (Kozai 1962).
Provided all other dynamical influences are negligible (see section 2),
an initially very small eccentricity in the inner binary will oscillate
through a maximum value given by
\begin{equation}
e_{1,{\rm max}}\simeq\left(1-{5\over3}\cos^2i_0\right)^{1/2},
\end{equation}
provided $|\cos i_0|<(3/5)^{1/2}$, i.e. the initial mutual inclination angle
$i_0$ lies between roughly $39^\circ$ and $141^\circ$ (Kozai 1962).
Given that the gravitational wave merger time is a strong function of
eccentricity (approximately proportional to $[1-e_1^2]^{7/2}$), highly
inclined orbits could in principle greatly reduce the merger time.  For
example, a mutual inclination angle of $56^\circ$ results in eccentricity
oscillations which, when at maximum amplitude, reduce the
characteristic merger time by an order of magnitude.  Although the true
merger time will depend on how long the inner binary spends at high
eccentricity, it appears promising that a random distribution of initial
inclinations will result in dramatically accelerated mergers in many cases.
This is the subject we will explore in the present paper.

While this research was being completed, we learned of recent work by Miller
\& Hamilton (2002), who investigate a very similar idea:  the use of the
Kozai mechanism to accelerate binary mergers in triple black hole
systems formed in globular clusters.

We begin in section 2 by going through the characteristic
time scales which will affect the dynamics of triple supermassive black
hole systems.  Then in section 3 we summarize the evolution equations of
an isolated triple used in our numerical calculations.  Section 4 presents
the results of those calculations, which help delineate the regions
of initial condition parameter space where the Kozai mechanism plays a
substantial role in accelerating the merger of the inner binary.  We discuss
our results and summarize our conclusions in section 5.

\section{CHARACTERISTIC TIME SCALES}

Consider a binary system consisting of two black holes with masses
$m_0$ and $m_1$, with initial semimajor axis $a_1$ and eccentricity $e_1$.
The time it takes for the binary to merge due to gravitational wave emission
(Peters 1964) can be written as
\begin{eqnarray}
\label{tmerge}
t_{\rm merge,binary} &\simeq& 2.9\times10^{12}{\rm yr}\left({m_0\over10^6\msun}
\right)^{-1}\left({m_1\over10^6\msun}\right)^{-1}\left({m_0+m_1\over
2\times10^6\msun}\right)^{-1}\left({a_1\over10^{-2}{\rm pc}}\right)^4
\nonumber \\
& & \times f(e_1)(1-e_1^2)^{7/2},
\end{eqnarray}
where $f(e_1)$ is a weak function of the initial eccentricity that is of
order unity ($0.979<f(e_1)<1.81$ for all $e_1$).
This gravitational wave merger time is a strong function of both the
semimajor axis and the eccentricity.  Hence gravitational radiation
only becomes important late in the evolution of the binary.

Prior to that time, the evolution is dominated by interactions between the
binary and surrounding material, either stars or gas.  These interactions
are very complex, due to the fact that the surroundings themselves evolve due
to their interaction with the binary.  (See Milosavljevi\'c \& Merritt 2001
for a recent discussion of stellar interactions.)  As we noted above, it
is not yet clear whether these interactions are sufficient to harden the
binary to a point where gravitational radiation then causes it to merge.

The scenario we wish to explore in this paper is one where the binary's
semimajor axis evolution has stalled because of insufficient interactions
with surrounding material, and where a subsequent galaxy merger then
introduces a third black hole in the
system.  We envisage the third black hole eventually forming a bound,
hierarchical triple with the binary.  The tidal gravitational torques exerted
on the inner binary by the outer black hole can alter
the eccentricity of the inner binary, thereby affecting the rate of
gravitational wave emission.

General relativity causes periastron precession in the inner binary, with
a period
\begin{equation}
\label{PGR}
P_{\rm GR}\simeq2.3\times10^6{\rm yr}\left({m_0+m_1\over2\times10^6\msun}
\right)^{-3/2}\left({a_1\over10^{-2}{\rm pc}}\right)^{5/2}(1-e_1^2).
\end{equation}
Provided this general relativistic precession
is unimportant, and the outer black hole is in a sufficiently inclined
orbit around the inner binary, then the eccentricity of the inner
binary will oscillate by the Kozai mechanism.  The characteristic time scale
for these oscillations
is given by (e.g. Holman, Touma, \& Tremaine 1997)
\begin{equation}
\label{Pe}
P_e\simeq1.3\times10^5{\rm yr}\left({m_0+m_1\over2\times10^6\msun}
\right)^{-1/2}\left({a_1\over10^{-2}{\rm pc}}\right)^{3/2}\left({m_0+m_1
\over2m_2}\right)\left({a_2/a_1\over10}\right)^3(1-e_2^2)^{3/2},
\end{equation}
where $m_2$ is the mass of the third body and $a_2$ and $e_2$ are the
semimajor axis and eccentricity, respectively, of its orbit around the
inner binary.

General relativistic precession can stop these
eccentricity oscillations by destroying the Kozai resonance (e.g. Holman et al.
1997).  A precise criterion for this not to happen is
easily derived (see equation [\ref{a2cubed}] in
Appendix):\footnote{Apart from the dependence on $e_1$
and factors of order unity, this is roughly equivalent to $P_e<P_{\rm GR}$.
Equation (\ref{a2a1}) provides a more quantitatively accurate description of the
behavior seen in our numerical results.}\/
\begin{equation}
\label{a2a1}
{a_2\over a_1}<34\left({a_1\over10^{-2}{\rm pc}}\right)^{1/3}
\left({m_0+m_1\over2\times10^6\msun}\right)^{-1/3}
\left({2m_2\over m_0+m_1}\right)^{1/3}\left({1-e_1^2\over1-e_2^2}\right)^{1/2}.
\end{equation}
The outer black hole must therefore come quite close to the inner binary
for eccentricity oscillations to take place, although the triple is
still hierarchical in the sense that the semimajor axis ratio $a_2/a_1$ is
large.

Interactions with the surrounding stars and gas may cause the outer black
hole's semimajor axis to evolve significantly over the merger time scale
of the inner binary, a point to which we shall return in section 5 below.
For now, however, we will neglect this fact and consider the evolution
of an isolated, hierarchical black hole triple.  As we will show, once the
outer black hole is sufficiently close that equation (\ref{a2a1}) is
satisfied, the merger time of the inner binary drops substantially for
sufficiently high mutual inclinations.  

\section{EQUATIONS OF MOTION OF AN ISOLATED TRIPLE SYSTEM}

We assume that the triple is hierarchical, so that the orbit of the outer
black hole is much larger in size than the orbit of the inner binary.  In this
case the triple can be considered to consist of two binaries in approximately
Keplerian orbits: the inner binary consisting of black holes with masses $m_0$
and $m_1$; and the outer binary consisting of the center of mass of the inner
binary, with mass $m_0+m_1$, and the third black hole with mass $m_2$.  Let
the semimajor axes of the inner and outer binaries be $a_1$ and $a_2$,
respectively, and the eccentricities be $e_1$ and $e_2$.  We also define
$g_1$ and $g_2$ to be the corresponding arguments of periastron.

The magnitude of the angular momenta ${\bf G}_1$ and ${\bf G}_2$ of the
inner and outer binaries are given by
\begin{equation}
G_1=m_0m_1\left[{Ga_1(1-e_1^2)\over m_0+m_1}\right]^{1/2}
\end{equation}
and
\begin{equation}
G_2=(m_0+m_1)m_2\left[{Ga_2(1-e_2^2)\over m_0+m_1+m_2}\right]^{1/2}
\end{equation}
respectively, where $G$ is Newton's gravitational constant.
Let the total orbital angular momentum of the triple be
${\bf H}={\bf G}_1+{\bf G}_2$.  In the absence of gravitational radiation,
tidal torques, or gravitational interactions with surrounding stars, this
vector would be rigorously conserved.  Let $i_1$ ($i_2$) be the inclination
of the inner (outer) binary, i.e. the angle between ${\bf G}_1$ (${\bf G}_2$)
and ${\bf H}$.  Then
\begin{equation}
H=G_1\cos i_1+G_2\cos i_2
\end{equation}
and
\begin{equation}
G_1\sin i_1 = G_2\sin i_2.
\end{equation}
The mutual inclination angle between the two binaries is $i=i_1+i_2$.

Marchal (1990), Krymolowski \& Mazeh (1999) and Ford, Kozinsky, \& Rasio
(2000) have derived the orbit-averaged Hamiltonian of an isolated, Newtonian
hierarchical triple of point masses, using secular perturbation theory to
octupole order, i.e. to order $(a_1/a_2)^3$.  We adopt the
equations of motion for the orbital elements of Ford et al.
here,\footnote{The equations of motion
of Krymolowski \& Mazeh (1999) differ from those of Ford et al. (2000) by
terms of order $(a_1/a_2)^{7/2}$ resulting from the canonical transformation
of the von Zeipel method.  We have
corrected a sign error in the
octupolar terms in the equations of motion of Ford et al. (2000).  In
equations (22) and (29)-(32) of their paper, all terms involving $C_3$
should have the opposite sign.  The equations of motion
of Krymolowski \& Mazeh (1999) also contain the same sign error.}\/
but modify them to incorporate two general
relativistic effects on the inner binary: the precession of periastron and
gravitational radiation.  We do this by simply adding orbit-averaged general
relativistic
correction terms to the Ford et al. (2000) equations of motion for the orbital
elements of the inner binary, the same correction terms which would exist if
the inner binary were isolated.  We must also take into account the fact
that ${\bf H}$ is no longer conserved because of the radiative loss
$(dG_1/dt)_{\rm rad}$ of the inner binary's orbital angular momentum.  We
do this by noting that in the absence of interactions between the inner
and outer binaries, gravitational radiation acts to change the magnitude of
${\bf G}_1$, but not its direction.  In addition, the vector ${\bf G}_2$
remains unchanged.  The resulting equation is
\begin{equation}
{dH\over dt}={G_1+G_2\cos i\over H}\left({dG_1\over dt}\right)_{\rm rad}.
\label{eqdhrad}
\end{equation}

Note that if the black holes are spinning, then the orbital plane of the
inner binary can also evolve because of general relativistic spin-orbit
coupling. This is a higher order effect than general relativistic precession,
and we neglect it entirely, assuming that we are essentially dealing with
Schwarzschild black holes.

Thus our evolution equations for the orbital elements are

\begin{equation}
{da_1\over dt}=-{64G^3m_0m_1(m_0+m_1)\over5c^5a_1^3(1-e_1^2)^{7/2}}
               \left(1+{73\over24}e_1^2+{37\over96}e_1^4\right),
\label{da1dt}
\end{equation}
\begin{eqnarray}
{dg_1\over dt}&=&6C_2\left\{{1\over G_1}[4\theta^2+(5\cos2g_1-1)(1-e_1^2-
                 \theta^2)]+{\theta\over G_2}[2+e_1^2(3-5\cos2g_1)]\right\}
                 \nonumber\\
              & &+C_3e_2e_1\left({1\over G_2}+{\theta\over G_1}\right)
                 \left\{\sin g_1\sin g_2[A+10(3\theta^2-1)(1-e_1^2)]-
                 5\theta B\cos\phi\right\}\nonumber\\
              & &-C_3e_2{1-e_1^2\over e_1G_1}\left[10\theta(1-\theta^2)
                 (1-3e_1^2)\sin g_1\sin g_2+\cos\phi(3A-10\theta^2+2)\right]
                 \nonumber\\
              & &+{3\over c^2a_1(1-e_1^2)}\left[{G(m_0+m_1)\over a_1}
                 \right]^{3/2},
\label{dg1dt}
\end{eqnarray}
\begin{eqnarray}
{de_1\over dt}&=&30C_2{e_1(1-e_1^2)\over G_1}(1-\theta^2)\sin2g_1\nonumber\\
              & &-C_3e_2{1-e_1^2\over G_1}\big[35\cos\phi(1-\theta^2)e_1^2
                 \sin2g_1-10\theta(1-e_1^2)(1-\theta^2)\cos g_1\sin g_2
                 \nonumber\\
              & &-A(\sin g_1\cos g_2-\theta\cos g_1\sin g_2)\big]\nonumber\\
              & &-{304G^3m_0m_1(m_0+m_1)e_1\over15c^5a_1^4(1-e_1^2)^{5/2}}
                 \left(1+{121\over304}e_1^2\right),
\label{de1dt}
\end{eqnarray}
\begin{equation}
{da_2\over dt}=0,
\end{equation}
\begin{eqnarray}
{dg_2\over dt}&=&3C_2\left\{{2\theta\over G_1}[2+e_1^2(3-5\cos2g_1)]+
                 {1\over G_2}[4+6e_1^2+(5\theta^2-3)(2+3e_1^2-5e_1^2
                 \cos2g_1)]\right\}\nonumber\\
              & &-C_3e_1\sin g_1\sin g_2\bigg\{{4e_2^2+1\over e_2G_2}10\theta
                 (1-\theta^2)(1-e_1^2) \nonumber\\
              & &\qquad\qquad\qquad\qquad\quad -e_2\left({1\over G_1}+
                 {\theta\over G_2}\right)[A+10(3\theta^2-1)(1-e_1^2)]\bigg\}
                 \nonumber\\
              & &-C_3e_1\cos\phi\left[5B\theta e_2\left({1\over G_1}+
                 {\theta\over G_2}\right)+{4e_2^2+1\over e_2G_2}A\right],
\end{eqnarray}
\begin{equation}
{de_2\over dt}=C_3e_1{1-e_2^2\over G_2}\left[10\theta(1-\theta^2)(1-e_1^2)
               \sin g_1\cos g_2+A(\cos g_1\sin g_2-\theta\sin g_1
               \cos g_2)\right],
\label{de2dt}
\end{equation}
and
\begin{equation}
{dH\over dt}=-{32G^3m_0^2m_1^2\over 5c^5a_1^3(1-e_1^2)^2}
             \left[G(m_0+m_1)\over a_1\right]^{1/2}
             \left(1+{7\over 8}e_1^2\right){G_1+G_2\theta\over H}.
\label{dhdt}
\end{equation}

The quantities $C_2$ and $C_3$ multiply the quadrupole and octupole
perturbation terms, respectively.  They are defined by (Ford et al. 2000)
\begin{equation}
C_2={Gm_0m_1m_2\over16(m_0+m_1)a_2(1-e_2^2)^{3/2}}\left({a_1\over a_2}
\right)^2,
\label{eqc2}
\end{equation}
and
\begin{equation}
C_3={15Gm_0m_1m_2(m_0-m_1)\over64(m_0+m_1)^2a_2(1-e_2^2)^{5/2}}
\left({a_1\over a_2}\right)^3.
\label{eqc3}
\end{equation}
Note that the octupole terms vanish if $m_0=m_1$.
The quantities $B$ and $A$ in these terms are given by
\begin{equation}
B=2+5e_1^2-7e_1^2\cos2g_1
\end{equation}
and
\begin{equation}
A=4+3e_1^2-{5\over2}(1-\theta^2)B.
\end{equation}
Finally, the quantity $\theta$ is the cosine of the mutual inclination of
the binaries,
\begin{equation}
\theta=\cos i={H^2-G_1^2-G_2^2\over2G_1G_2},
\label{eqtheta}
\end{equation}
and $\phi$ is the angle between the periastron directions,
\begin{equation}
\cos\phi=-\cos g_1\cos g_2-\theta\sin g_1\sin g_2.
\end{equation}
Note that equation (\ref{eqtheta}) determines the evolution of the
mutual inclination through the time dependence of the eccentricities, $a_1$,
and $H$.
The terms in the equations of motion which remain after setting $C_2=C_3=0$
are the general relativistic correction terms.
The last term in equation (\ref{dg1dt}) is the general relativistic precession
term, and the terms involving the speed of light $c$ in equations (\ref{da1dt}),
(\ref{de1dt}), and (\ref{dhdt}) are the gravitational radiation terms.

We may immediately deduce some important features of our evolution equations.
Equations (\ref{de1dt}) and (\ref{de2dt}) imply that if both the
inner and outer binaries are circular ($e_1=e_2=0$), then they will remain
that way.  An eccentric outer binary will produce a nonzero eccentricity
in an initially circular inner binary, and vice-versa, but only because of
the octupole interaction terms.

Note that if we switch off the quadrupole and octupole interaction terms
by setting $C_2=C_3=0$, then an immediate
consequence is $d\theta/dt=0$, as may be verified by differentiating equation
(\ref{eqtheta}) directly.  In other words, general relativistic effects alone
do not change the mutual inclination of the inner and outer binaries if there
are no interactions between them.  This is as one would expect, given
our treatment of the gravitational wave angular momentum loss 
in equation (\ref{eqdhrad}).

Our evolution equations also exhibit an important scaling.  If all masses
and all initial semimajor axes are multiplied by some constant factor, then
the merger time and all other time scales change by the same multiplicative
factor.  This is a direct consequence of the nature of gravity, where mass,
length, and time have the same dimensions in units where $G=c=1$, and it
reduces the parameter space of masses and semimajor axes which need to
be explored numerically.  Another similarly useful fact is that the equations
are unchanged when $m_0$ and $m_1$ in the inner binary are interchanged
and $g_1$ or $g_2$ is increased by $180^\circ$.  The former corresponds to
a relabeling of the inner binary black holes, while the latter represents
a spatial inversion of the two orbits.

While our implementation of the general relativistic effects has
been heuristic, the general relativistic precession terms for the inner
binary can in fact be rigorously justified by orbit-averaging the
post-Newtonian Hamiltonian.  To octupole order, the resulting
Hamiltonian is
\begin{eqnarray}
{\bar{\cal H}(g_1,g_2,G_1,G_2)}&=&C_2\left[(2+3e_1^2)(1-3\theta^2)-15e_1^2
(1-\theta^2)\cos2g_1\right]
\nonumber\\
& &+C_3e_1e_2\left[A\cos\phi+10\theta(1-\theta^2)(1-e_1^2)
\sin g_1\sin g_2\right]
\nonumber\\
& &+{G^2m_0m_1\over c^2a_1^2}\left[{15m_1^2+15m_0^2+29m_0m_1\over8(m_0+m_1)}
-{3(m_0+m_1)\over(1-e_1^2)^{1/2}}\right].
\label{hamav}
\end{eqnarray}
A rigorous derivation of the radiation terms
would be considerably more difficult, and is beyond the scope of this
paper.\footnote{One of the issues is that the Newtonian portion of the
equations of motion were derived by using the invariable plane as the
reference plane.
Because this plane slowly evolves under the influence of gravitational
radiation, it may be that a rigorous set of equations of motion
should include terms to reflect this slow evolution.}
We believe that our equations of motion
will capture the overall time evolution of the inner binary, because the early
phases of this evolution will be dominated by interactions with the outer
black hole, while the late phases will be dominated by gravitational
radiation.
In both those limits our evolution equations for the orbital elements of the
inner binary are rigorously correct.  It is possible, however, that
interesting
effects may occur during the transition between these two phases of evolution,
which may not be captured by our equations.

In the absence of gravitational radiation, our equations conserve
the orbit averaged Hamiltonian in equation (\ref{hamav}).
We use this fact to check the accuracy of our numerical integrations.

The reader should bear in mind that our equations are based on orbit
averaging and an expansion in the semimajor axis ratio $a_1/a_2$, and may
therefore be inaccurate when
this ratio is not very small.  We will show in the next section that the
octupole terms usually have fairly small effects on the merger time of the
inner binary.  Hence our neglect of even higher order terms, and our use
of such an expansion in the first place, is probably justified for the
primary purpose of this paper.

The major exception to this is the issue of stability.  We are investigating
whether the presence of a third black hole can cause the inner binary to
merge before the third black hole comes close enough to cause an
unstable interaction.  There is a limit to how small $a_2/a_1$ can be for
the triple to be stable and for the evolution we calculate to be valid.
Completely general stability criteria for hierarchical triples in arbitrary
initial configurations do not exist.  We adopt as a stability criterion the
empirical formula used by Mardling \& Aarseth (2001), which may be written as
\begin{equation}
{a_2\over a_1}>{2.8\over1-e_2}\left[\left(1+{m_2\over m_0+m_1}\right){1+e_2
\over(1-e_2)^{1/2}}\right]^{2/5}.
\label{eqacrit}
\end{equation}
This criterion was derived for Newtonian coplanar prograde orbits of the
inner and
outer binaries.  Inclined orbits, which are the major focus of this paper,
are expected to be more stable, so inequality (\ref{eqacrit}) provides
a conservative stability limit.  This inequality is also conservative
in the sense that it is either close to or above the more
complicated stability criterion proposed by Eggleton \& Kiseleva (1995),
which was empirically verified over a wide
range of mass ratios including nearly all those we numerically investigate in
the next section.  All the calculations we present in this paper are done
for systems which are stable according to inequality (\ref{eqacrit}).

The stability criterion (\ref{eqacrit}) may be combined with the constraint
of inequality (\ref{a2a1}) that general relativistic precession not destroy
the Kozai resonance to give a lower limit on the inner semimajor axis for
which the Kozai mechanism can reduce the inner binary merger time.
\footnote{We are grateful to the referee for suggesting this connection
between stability and the precession constraint, and for emphasizing to
us the importance of discussing the stability of triple systems in general.}
The result is
\begin{equation}
a_1\ga 6\times10^{-6}{\rm pc}\left({m_0+m_1+m_2\over m_0+m_1}\right)^{6/5}
\left({m_0+m_1\over2m_2}\right)\left({m_0+m_1\over2\times10^6{\rm M_\odot}}
\right){(1+e_2)^{27/10}\over(1-e_2)^{21/10}(1-e_1^2)^{3/2}}.
\label{a1crit}
\end{equation}
A stable triple with initial conditions such that the inner semimajor axis
$a_1$ violates this inequality will not have an accelerated inner binary
merger by the Kozai mechanism.  However, in that case $a_1$ is so small that
the binary merger time by equation (\ref{tmerge}) is
\begin{eqnarray}
t_{\rm merge, binary}&\la&0.3{\rm yr}\left({m_0\over10^6{\rm M_\odot}}
\right)^{-1}\left({m_1\over10^6{\rm M_\odot}}\right)^{-1}\left({m_0+m_1\over
2\times10^6{\rm M_\odot}}\right)^3\left({m_0+m_1+m_2\over m_0+m_1}
\right)^{24/5}\nonumber\\
&&\left({m_0+m_1\over2m_2}\right)^4{(1+e_2)^{54/5}f(e_1)\over
(1-e_2)^{42/5}(1-e_1^2)^{5/2}}.
\label{tmcrit}
\end{eqnarray}
At least for roughly equal mass triples and outer eccentricities that are
not too high, this implies that the inner binary would already be rapidly
merging and needs no help from the Kozai mechanism.

On the other hand, equations (\ref{a1crit}) and (\ref{tmcrit}) have a rather
strong dependence
on $e_2$, so if the outer black hole's orbit is {\it very}
eccentric, then the resulting triple may be too unstable for there to be
time for the Kozai mechanism to operate.
However, numerical simulations of black hole binary evolution by
Milosavljevi\'c
\& Merritt (2001) suggest that this is rather unlikely.  They find rather
modest initial eccentricities when the binary
first forms, and the subsequent evolution of the binary does not go to
very high eccentricity.
Hence there is probably plenty of parameter space
for stable hierarchical triples to form and evolve under the Kozai
mechanism.

\section{NUMERICAL RESULTS}

\subsection{Detailed Evolution of a Triple Consisting of Nearly Equal Mass
Black Holes}

We have numerically integrated equations (\ref{da1dt})-(\ref{dhdt}) for a
wide range of possible black hole masses and initial conditions.  Even for
fixed black hole masses, the parameter space of possible initial conditions
is huge, and it helps to understand which are the most important quantities
affecting the merger time.  We therefore focus first on the detailed
evolution of a particular triple consisting of nearly equal mass black
holes.  Specifically, we consider an inner binary with a
$2\times10^6$~M$_\odot$ black hole and a $10^6$~M$_\odot$ black hole, about
which orbits another $10^6$~M$_\odot$ black hole ten times further
out.  We have purposely chosen the two inner black holes to have different
masses so as to allow for octupole interaction effects in the evolution.

\begin{figure}[hb]
\plottwo{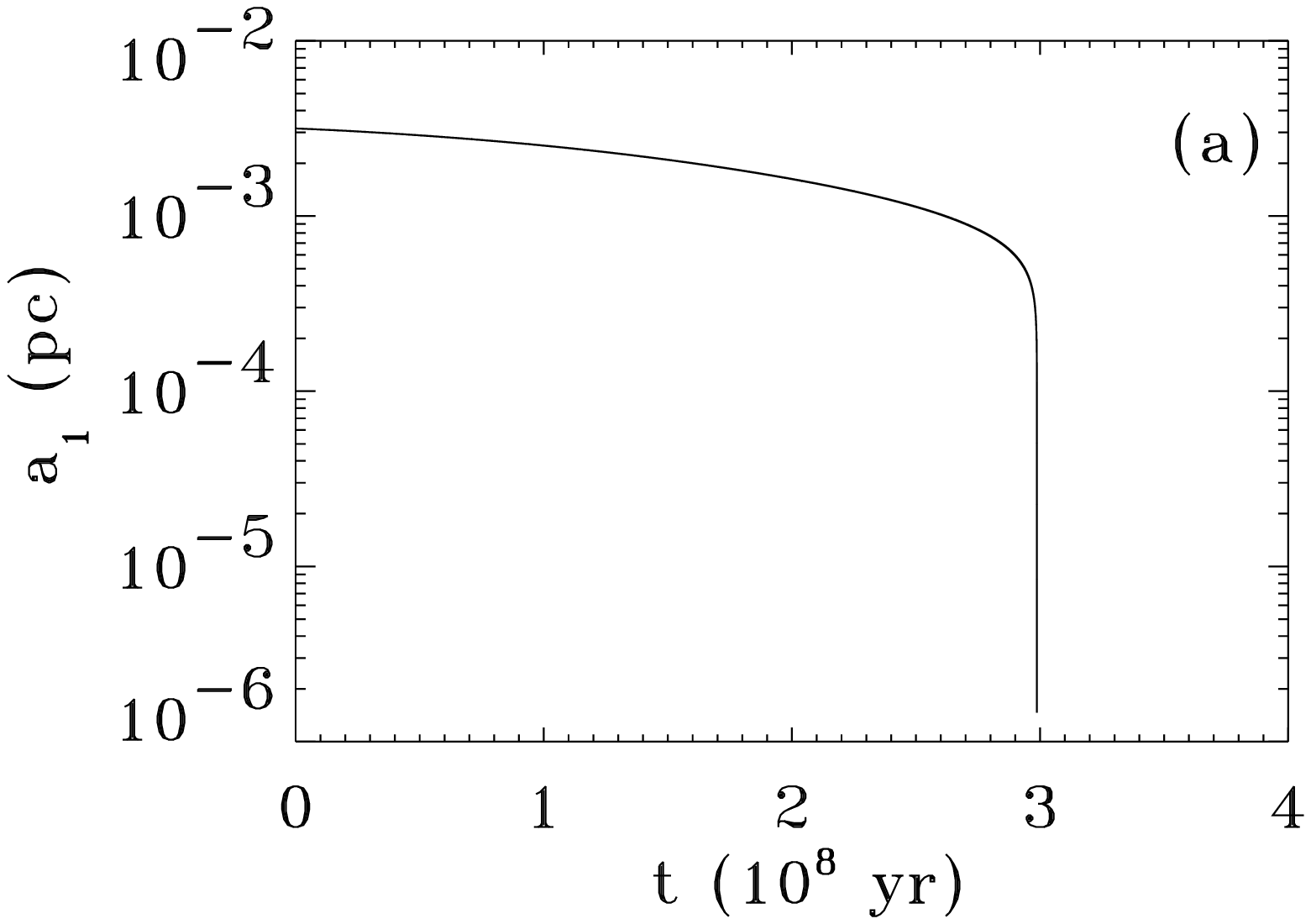}{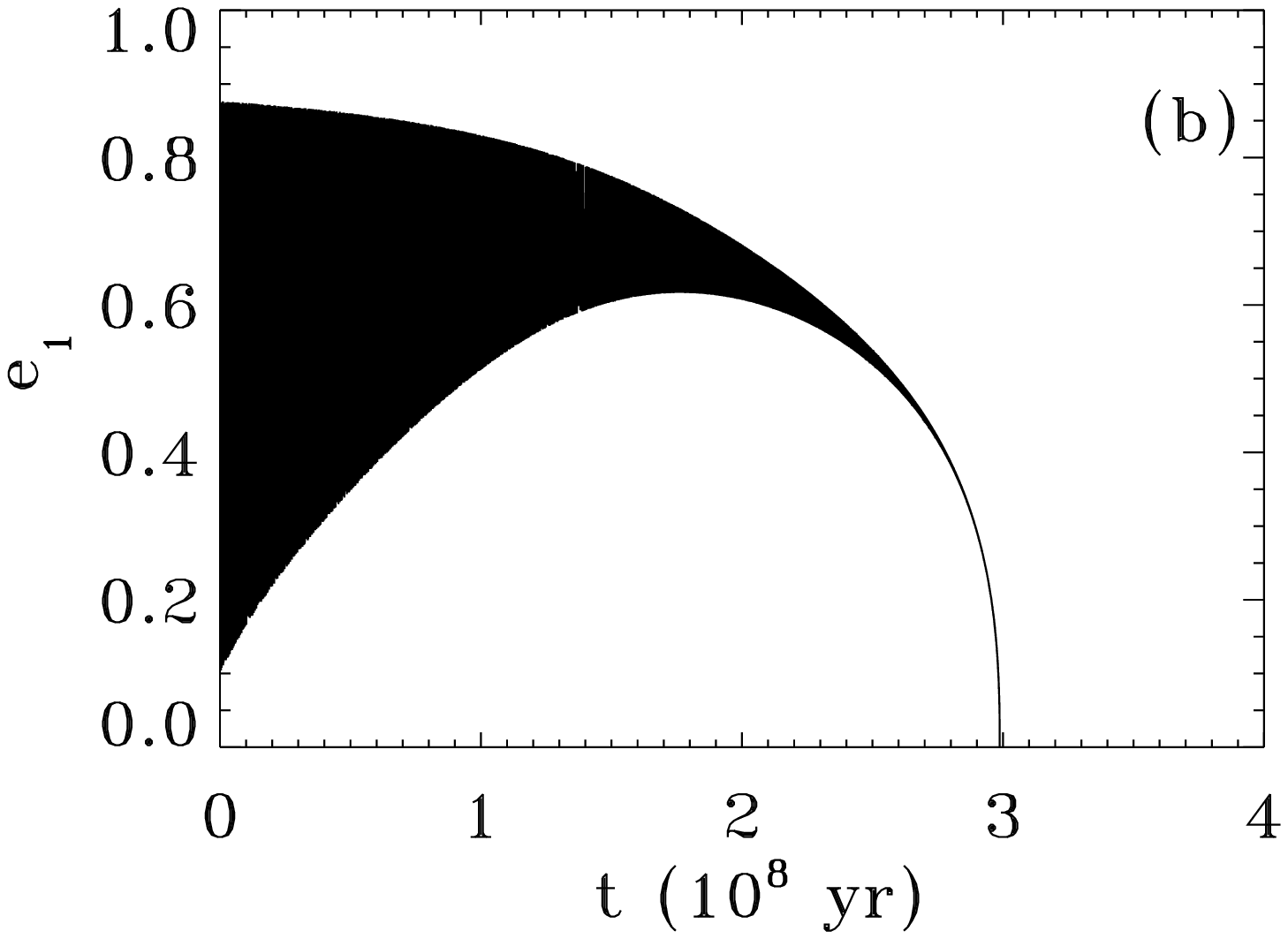}
\plotone{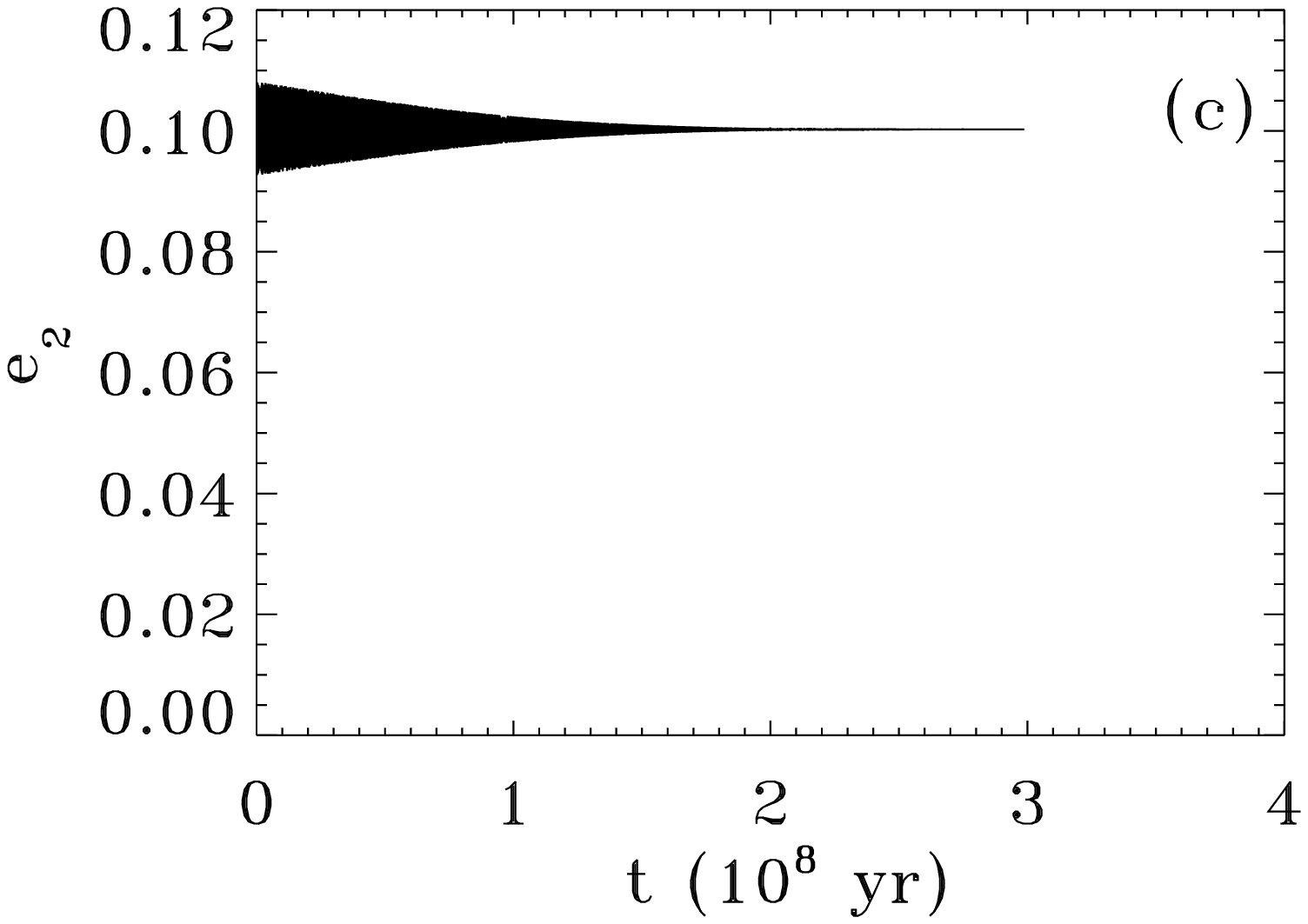}
\caption{Evolution of (a) the inner binary semimajor axis $a_1$, (b) the
inner binary eccentricity $e_1$, and (c) the outer binary eccentricity $e_2$
with time
for a triple consisting of black hole masses $m_0=2\times10^6$~M$_\odot$
and $m_1=m_2=10^6$~M$_\odot$.  The initial conditions of the
triple are $a_1=3.16\times10^{-3}$~pc, $a_2/a_1=10$, $e_1=0.1$, $e_2=0.1$,
$g_1=0$, $g_2=90^\circ$, and $i=80^\circ$.  The large amplitude inner binary
eccentricity oscillations due to the Kozai mechanism are clearly evident.
This greatly
accelerates the merger of the inner binary: in the absence of eccentricity
oscillations, the binary would take $9.3\times10^9$~yr to merge.}
\end{figure}

Figure 1 depicts the evolution of inner semimajor axis $a_1$, inner
eccentricity $e_1$, and outer eccentricity $e_2$ for a particular choice of
initial conditions.  The inner binary starts out nearly circular ($e_1=0.1$)
with a semimajor axis which would give a gravitational wave merger time
$t_{\rm merge,binary}=9.3\times10^9$~yr if the binary were isolated.  However,
the presence of the outer black hole induces large amplitude
eccentricity oscillations in the inner binary through the Kozai mechanism.
The time spent at higher eccentricity reduces the gravitational wave merger
time of the two inner black holes by roughly a factor thirty for the
particular case shown.

In contrast to $e_1$, the outer eccentricity $e_2$ stays close to its
original value throughout the evolution.  Equation (\ref{de2dt}) shows that
$e_2$ only changes as a result of octupole interactions, which are
weaker than the quadrupolar interactions driving the inner binary eccentricity
oscillations.  The lack of strong evolution in the outer eccentricity
is a generic feature of all the numerical calculations we have done,
implying that $e_2$ is unlikely to be driven to high enough values to
destabilize an isolated triple.

\begin{figure}[t]
\plottwo{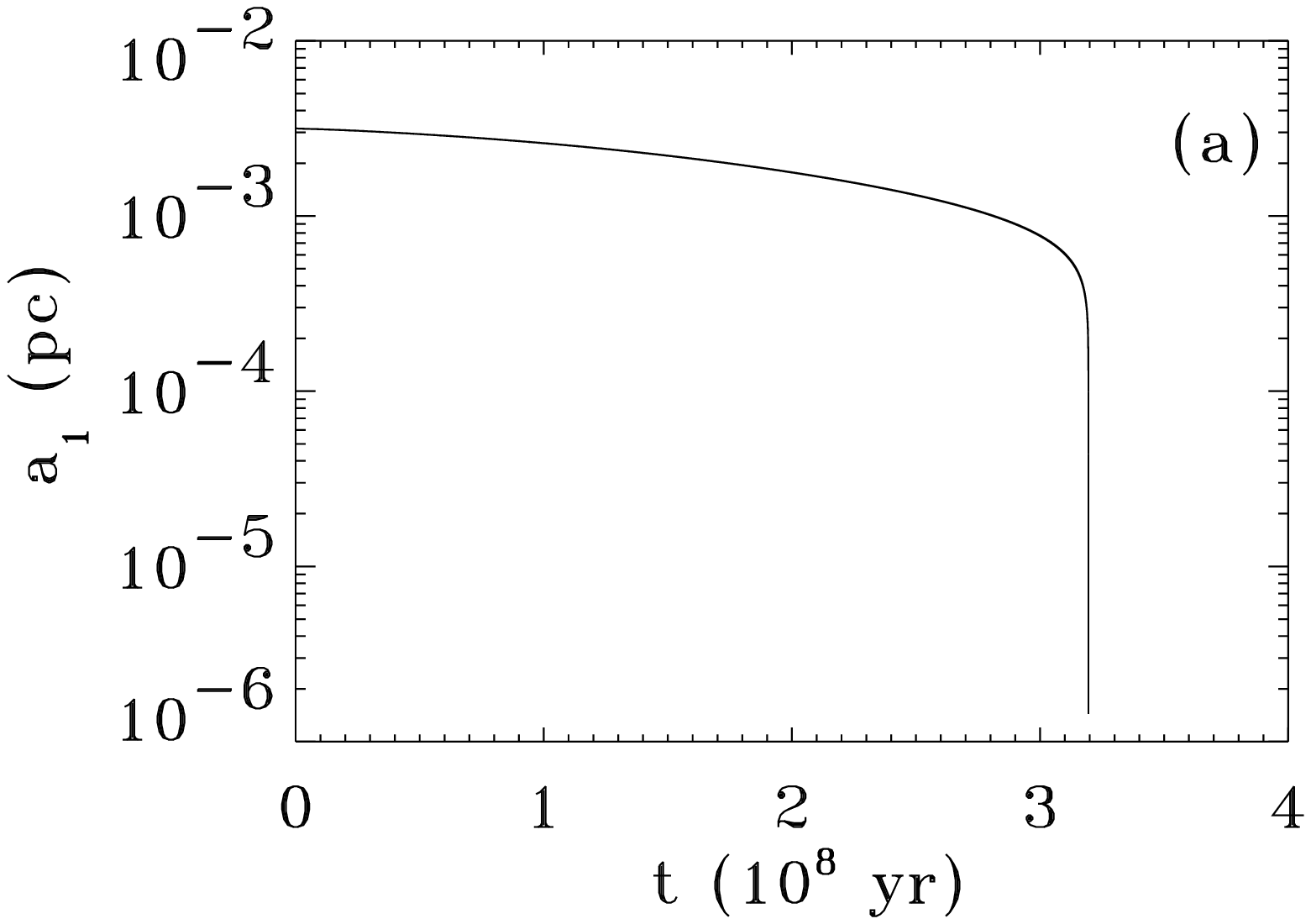}{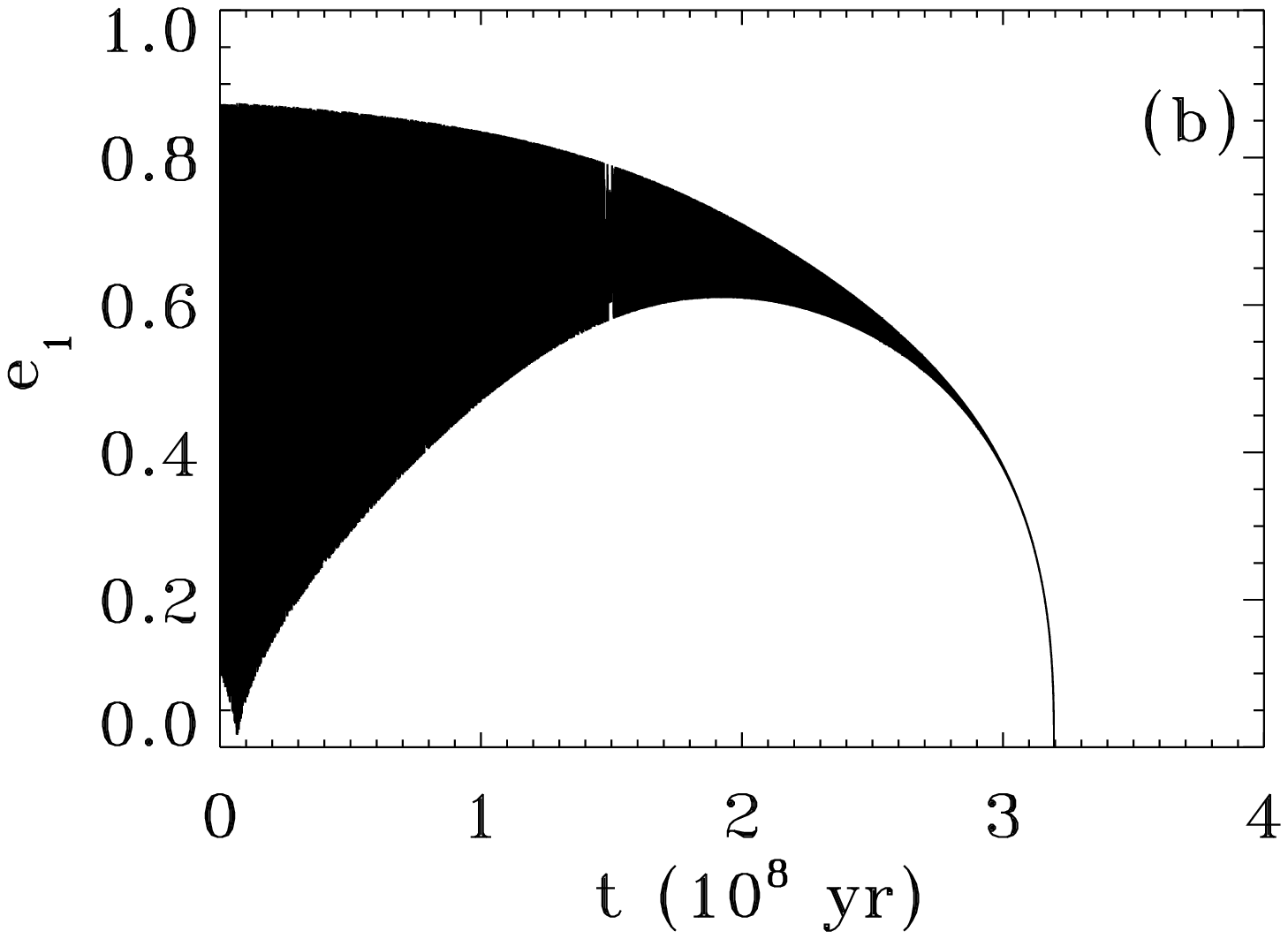}
\plotone{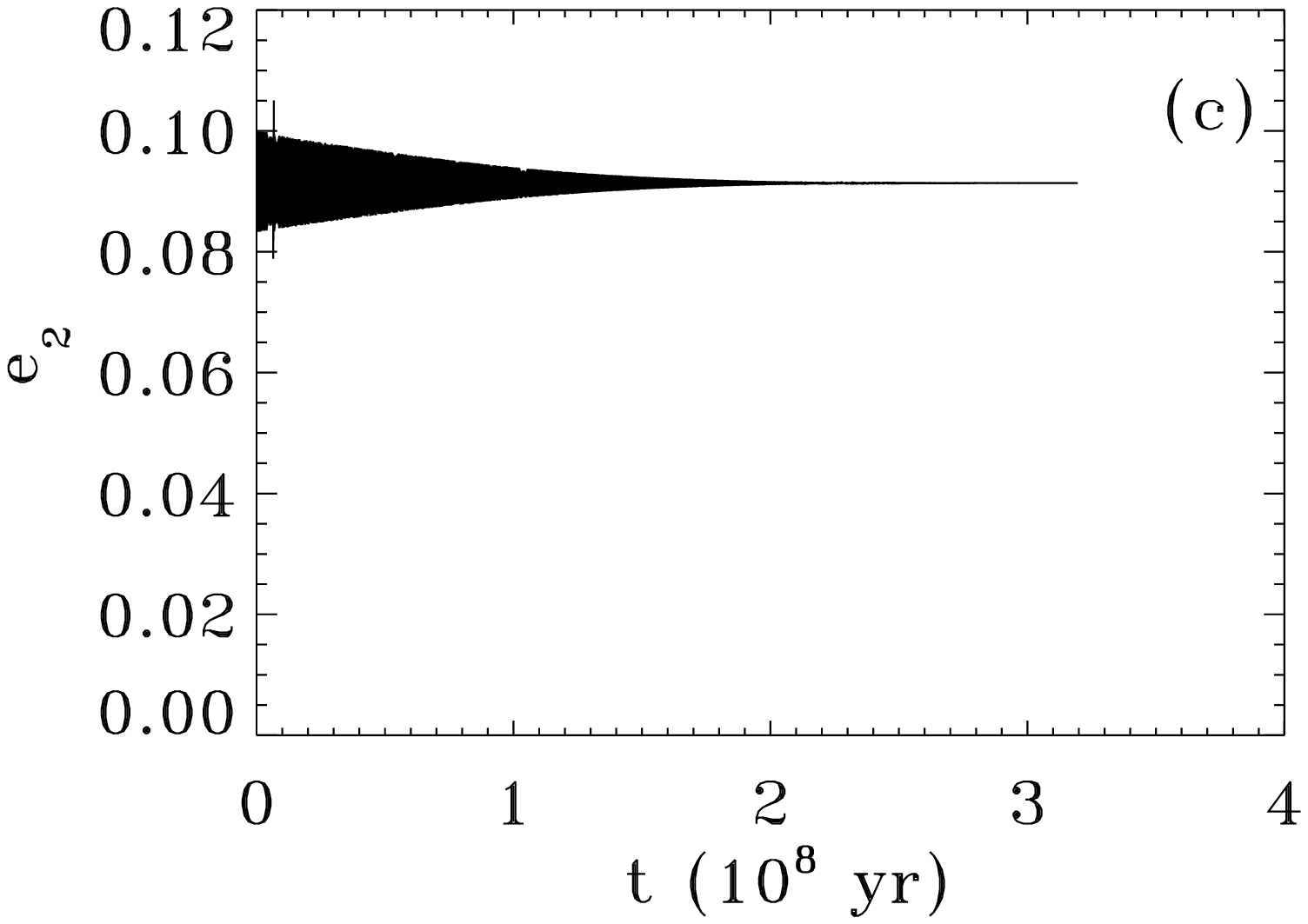}
\caption{Same as figure 1 except for an initial inner argument of periastron
$g_1=90^\circ$.}
\end{figure}

As might be expected, the merger time turns out to be rather insensitive to
the initial orientation of the roughly circular inner orbit, i.e. the initial
value of the inner argument of periastron $g_1$.  Figure 2 shows the evolution
of the same triple as depicted in figure 1, except that the initial value of
$g_1$ has been shifted by $90^\circ$.  This results in a merger time which
is about seven percent longer than that shown in figure 1.

\begin{figure}[t]
\epsscale{0.6}
\plotone{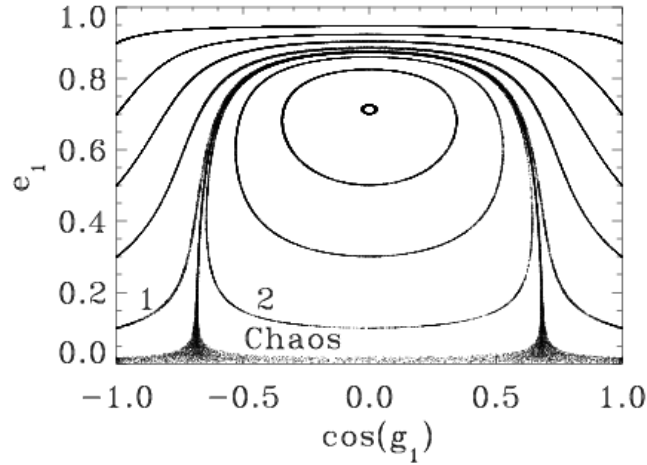}
\caption{Trajectories in the $e_1$ vs. $\cos g_1$ phase space for triples
with similar initial conditions to the triples in figures 1 and 2,
neglecting gravitational radiation.  Each curve
corresponds to a different initial inner eccentricity $e_1$ and
mutual inclination angle $i$ chosen to keep the total angular momentum of
the triple fixed.  The initial inner argument of periastron $g_1$ is also
chosen to be either 0 or $90^\circ$.  Provided the initial $e_1$
is not too small, the former choice always produces circulation in $g_1$, while
the latter can produce libration in $g_1$ about $90^\circ$ instead.  The
circulating curve labeled 1 and the librating curve labeled 2 have identical
initial conditions to the triples shown in figures 1 and 2, respectively.
Triples which start out with nearly circular inner binaries ($e_1$ small)
are chaotic, repeatedly crossing the separatrix between circulation and
libration.}
\end{figure}

The overall behavior of the eccentricity oscillations depicted in figures
1(b) and 2(b) can be understood by examining the evolution in the
$e_1$ vs. $g_1$ phase space.  The initial phase space trajectories for
triples with the same total angular momentum as the triples of figures
1 and 2 are depicted in figure 3.  Depending on the initial conditions,
the evolution in $g_1$ is either one of libration about $g_1=90^\circ$
or $270^\circ$, or one of circulation.  The quadrupolar fixed point of the
Kozai resonance lies inside the smallest libration contour at $\cos g_1=0$.
The only difference between the triples of figures
1 and 2 is that the former starts off circulating while the latter starts
off librating.  Figure 4 depicts snapshots of the phase space trajectories
during the course of the evolution shown in figures 1 and 2.  As shown
in figure 4(b), gravitational radiation drives an initially librating 
inner binary over into a circulating inner binary, thereby causing the
minimum eccentricity to drop until it crosses the separatrix.  This separatrix
crossing corresponds to the momentary zero eccentricity spike just before
$10^7$~years in figure 2(b).   Thereafter, the evolution is very similar
to the case where the inner binary starts off circulating:  the minimum
eccentricity rises until gravitational radiation becomes so strong that
it starts to circularize the orbit.

\begin{figure}[h]
\epsscale{1.0}
\plottwo{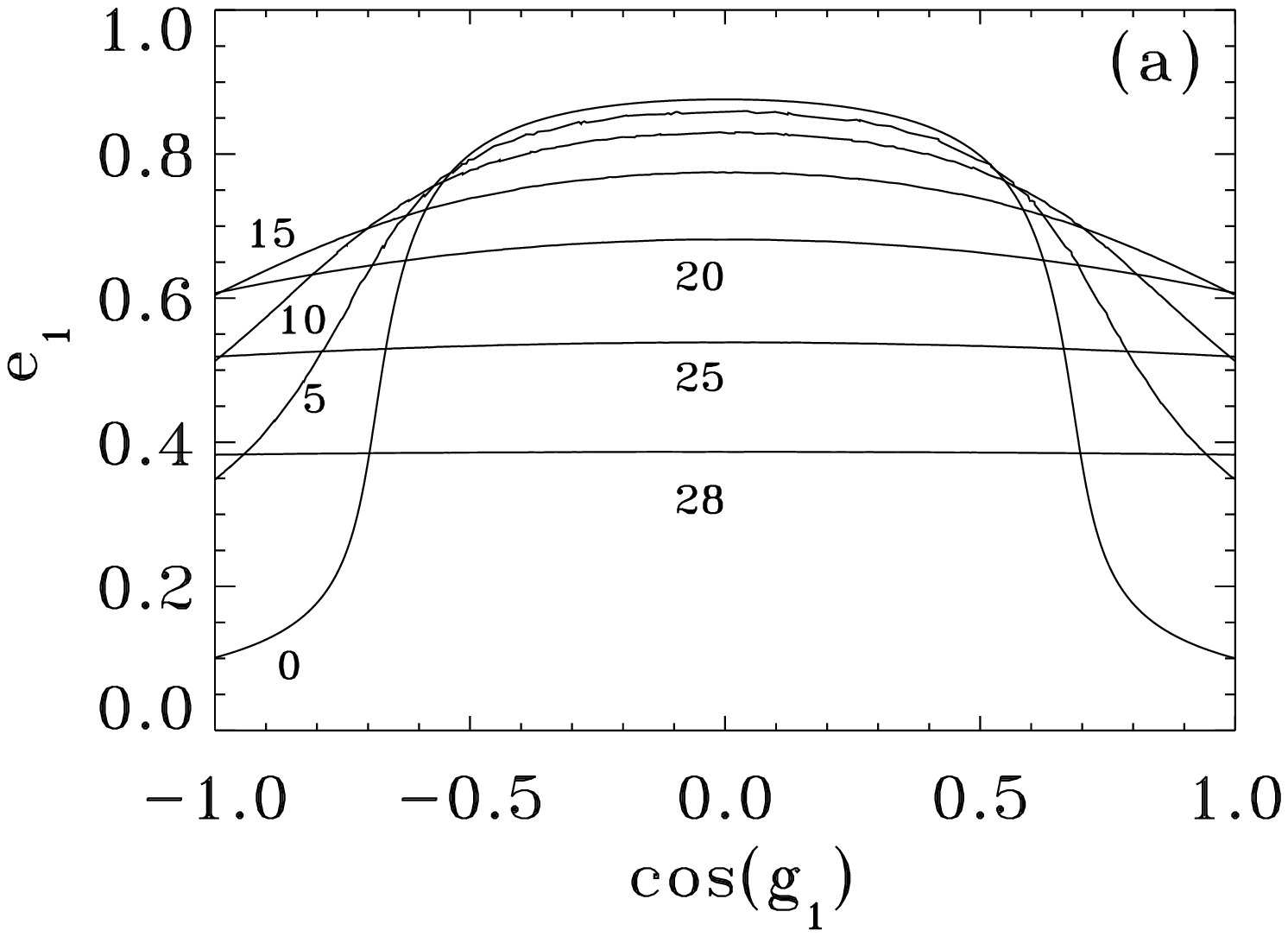}{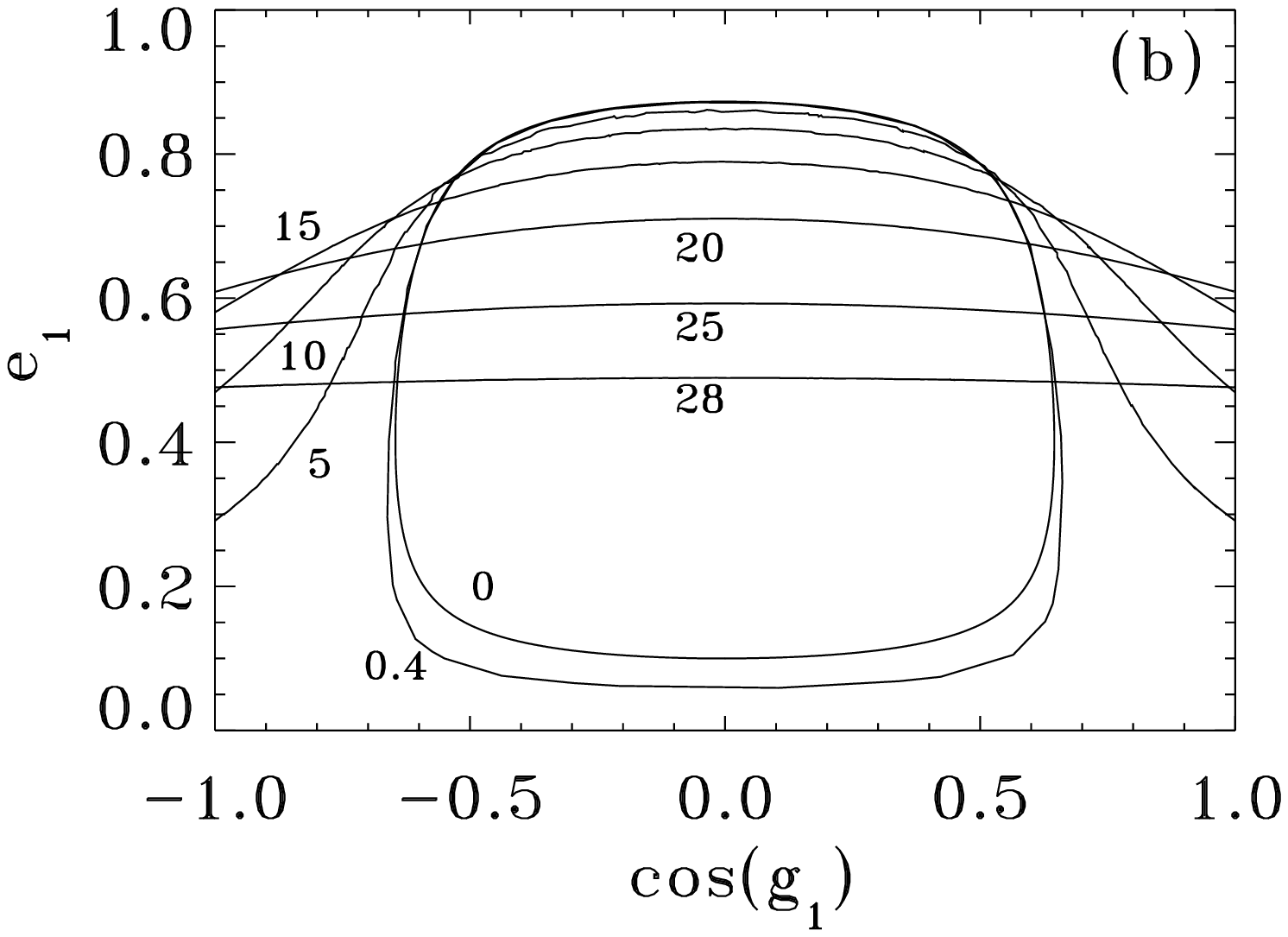}
\caption{Snapshots of the $e_1$ vs. $\cos g_1$ phase space during the course
of the evolution depicted in (a) figure 1 (initially circulating in $g_1$)
and (b) figure 2 (initially librating in $g_1$).  Each curve is labeled with
the time in units of $10^7$~yr.}
\epsscale{1.18}
\plotone{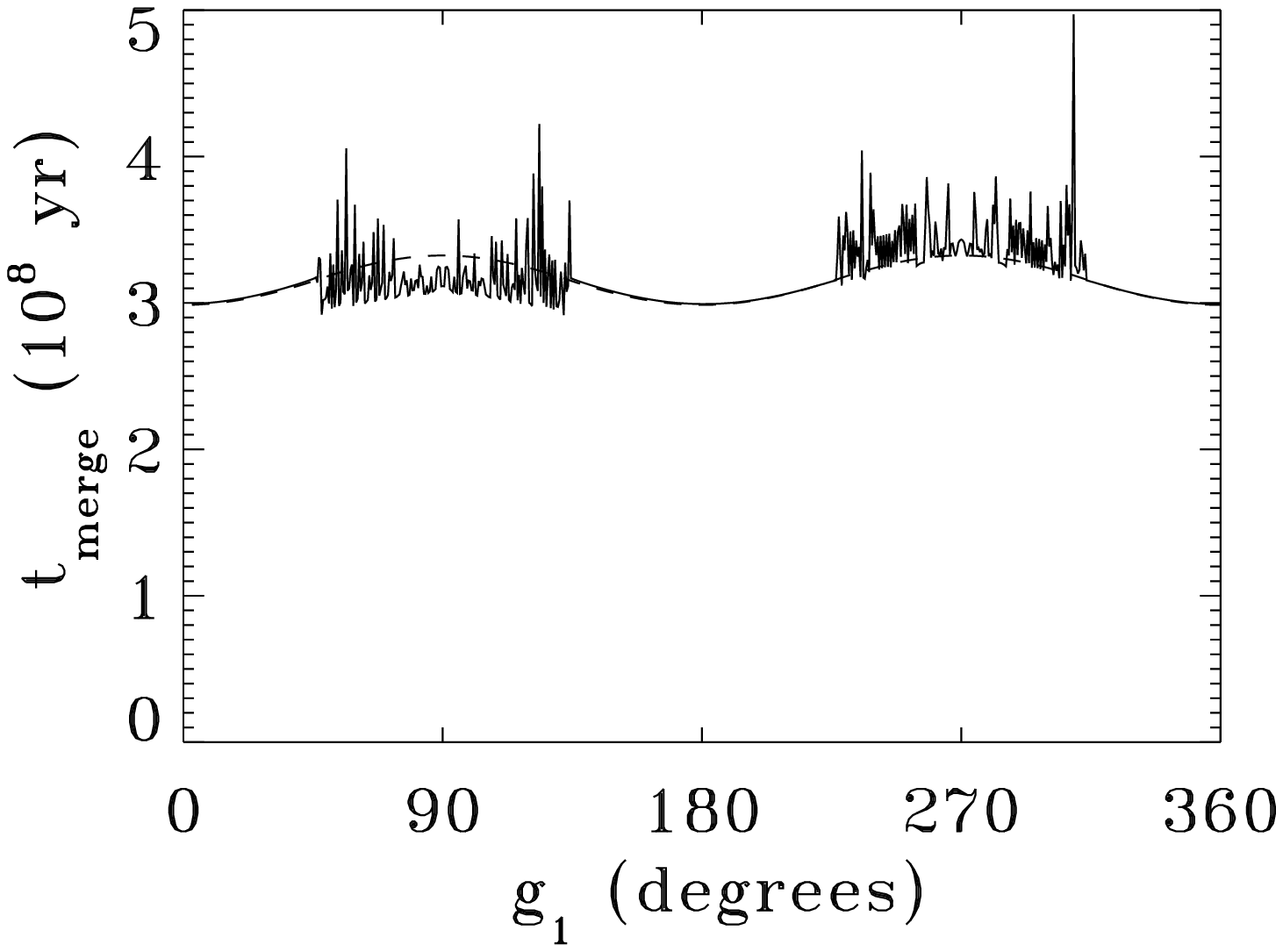}
\caption{Merger time as a function of the initial argument of periastron $g_1$
of the inner binary, for a triple consisting of black hole masses
$m_0=2\times10^6$~M$_\odot$ and $m_1=m_2=10^6$~M$_\odot$.  The initial
conditions of the triple are $a_1=3.16\times10^{-3}$~pc, $a_2/a_1=10$,
$e_1=0.1$, $e_2=0.1$, $g_2=90^\circ$, and $i=80^\circ$.  The solid
line depicts the merger time as calculated with our full equations of motion,
while the dashed curve is the time obtained by neglecting the octupole terms.}
\end{figure}

Figure 5 shows the overall dependence of the merger time for varying initial
inner argument of periastron $g_1$, again for an inner orbit which is initially
roughly circular: $e_1=0.1$.  While the overall effect of $g_1$ on
the merger time is not too great, the exact value of the merger time
is extremely sensitive to the initial value of $g_1$ for triples close
to the libration/circulation separatrix, or for those triples which
start out librating ($g_1$ initially around $90^\circ$ or $270^\circ$) and
therefore subsequently cross the separatrix.  As shown in figure 3, the
region near the separatrix is chaotic, and this chaos arises from the
octupole interaction terms.  If these terms are neglected,
the system then lacks the necessary degrees of freedom to exhibit chaos, and the
dependence of the merger time on the initial value of $g_1$ is much smoother
(the dashed line in figure 5).  Note that the separatrix is associated
with passing through very small values of $e_1$ [cf. figure 2(b)].  One might
therefore worry that our results suffer from numerical problems associated
with the $1/e_1$ singularity in the octupole term of equation (\ref{dg1dt}).
However, following the suggestion of Ford et al. (2000), we have removed this
singularity in our numerical integrations by first transforming the dependent
variables from $(g_1,e_1,g_2,e_2)$ to
$(e_1\cos g_1, e_1\sin g_1, e_2\cos g_2, e_2\sin g_2)$.  In addition, the
evolution of the Hamiltonian, equation (\ref{hamav}), is completely
smooth through the separatrix crossing, changing only as a result of
gravitational radiation losses.  We therefore believe the ``noise''
exhibited in figure 5 is physical, and is caused by chaotic behavior during
evolution through the separatrix.

\begin{figure}[t]
\epsscale{0.6}
\plotone{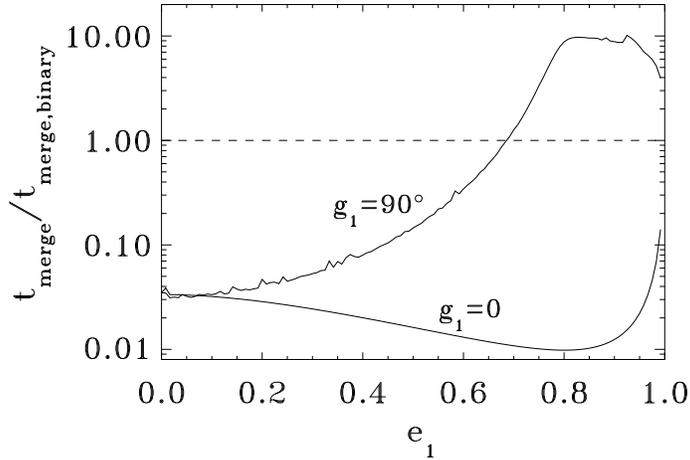}
\caption{Merger time as a function of the initial eccentricity $e_1$ of the
inner binary, for a triple consisting of black hole masses 
$m_0=2\times10^6$~M$_\odot$ and $m_1=m_2=10^6$~M$_\odot$.  The initial
conditions of the triple are $a_1=3.16\times10^{-3}$~pc, $a_2/a_1=10$,
$e_2=0.1$, $g_2=90^\circ$, and $i=80^\circ$.  Two initial values of
$g_1$ are shown: $g_1=0$ (initially circulating) and $g_1=90^\circ$
(initially librating for low $e_1$, circulating for high $e_1$).
In contrast to
previous figures, we have scaled the merger time with the nominal binary
merger time calculated neglecting interactions with the outer black hole,
equation (\ref{tmerge}), which also depends sensitively on $e_1$.
The Kozai mechanism speeds up the merger in all cases where the initial
eccentricity is low, and can even speed up the merger at high initial
eccentricity in the case of $g_1=0$.}
\end{figure}

The initial orientation of the inner orbit affects the merger time much more
substantially when the inner orbit is more eccentric.  Figure 6 depicts the
merger time as a function of the initial inner eccentricity $e_1$ for two
values of the initial inner argument of periastron: $g_1=0$ and $90^\circ$.
The former case corresponds to a circulating inner binary which
starts out with an eccentricity which is at the minimum in the Kozai oscillation
(see figure 3).  Hence in this case the Kozai mechanism {\it always} speeds up
the merger compared to the time $t_{\rm merge,binary}$ it would take an
isolated binary with the same initial eccentricity to merge.  In contrast,
inner binaries with initial $g_1=90^\circ$ start at the minimum in the
Kozai eccentricity oscillation for values of $e_1$ which are below the
Kozai fixed point, and at the maximum for values of $e_1$ above it.  We would
therefore expect the Kozai mechanism to reduce the merger time in the former
case and increase it in the latter.  This expectation is confirmed by the
behavior shown in figure 6, provided the binary does not start out too near
the Kozai fixed point which occurs at $e_1\simeq0.8$ for this set of triples
with fixed mutual inclination angles.  Near the Kozai fixed point, the merger
time is generally increased by the Kozai mechanism, due perhaps
to the fact that gravitational radiation drives the eccentricity at the Kozai
fixed point down to lower values with time.
We therefore conclude that the Kozai mechanism always
acts to reduce the merger time of an initially nearly circular binary, but
for eccentric inner binaries it will either reduce or increase the merger
time, depending on the orientation of the inner binary within the triple
system.

The $g_1=90^\circ$ curve in figure 6 exhibits the chaos we typically
find for inner binaries which start out librating, or circulating near the
separatrix.  This chaos is also present at the lowest values of eccentricity
$e_1$ in the $g_1=0$ curve, as the inner binary then starts out in the
chaotic zone shown in figure 3.

\begin{figure}[t]
\epsscale{0.6}
\plotone{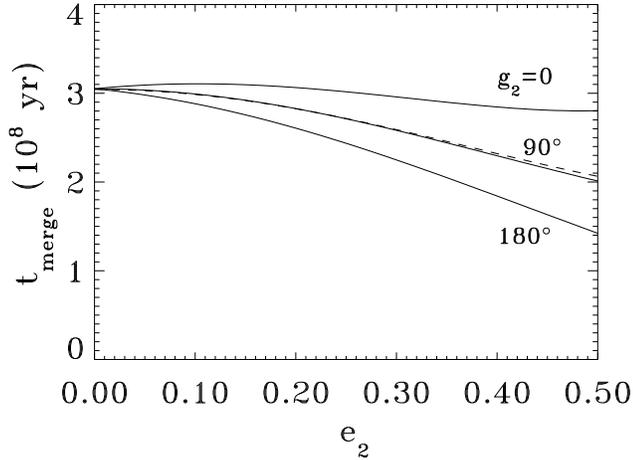}
\caption{Merger time as a function of the initial eccentricity $e_2$ of
the outer binary, for a triple
consisting of black hole masses $m_0=2\times10^6$~M$_\odot$ and
$m_1=m_2=10^6$~M$_\odot$.  The initial conditions of the triple are
$a_1=3.16\times10^{-3}$~pc, $a_2/a_1=10$, $g_1=0$, and $i=80^\circ$.
Different solid curves correspond to different choices for the initial
argument of periastron $g_2$ of the outer binary.  The dashed line shows
the merger time when octupole terms in the evolution equations are
neglected, in which case $g_2$ does not affect the evolution of the inner
binary.}
\end{figure}

We turn now to the effects of the orientation and eccentricity of the outer
binary on the merger time of the inner binary.
Equations (\ref{dg1dt})-(\ref{de1dt}) and (\ref{eqc2})-(\ref{eqc3}) show
that increasing the eccentricity $e_2$ of the outer binary at fixed
semimajor axis $a_2$ strengthens both the quadrupolar and octupolar
interaction terms, with the latter being enhanced over the former.  This
is of course physically reasonable as the distance of closest approach
of the outer black hole with the inner binary is smaller.  On the other
hand, the outer argument of periastron $g_2$ only affects the evolution of
the inner binary through the octupolar interaction terms.  We would therefore
expect that increasing $e_2$ would generally decrease the merger time
through the (quadrupolar) Kozai mechanism, but that the initial value of
$g_2$ could modify this significantly at high eccentricity due to the
enhanced octupolar effects.  These expectations are confirmed by our
numerical calculations shown in figure 7.  Note, however, that the effects
of $e_2$ and $g_2$ are not that large, at least for this particular
triple.  We have therefore chosen to fix their initial values to be
$e_2=0.1$ and $g_2=90^\circ$ for all our other numerical calculations.

\begin{figure}[t]
\epsscale{0.6}
\plotone{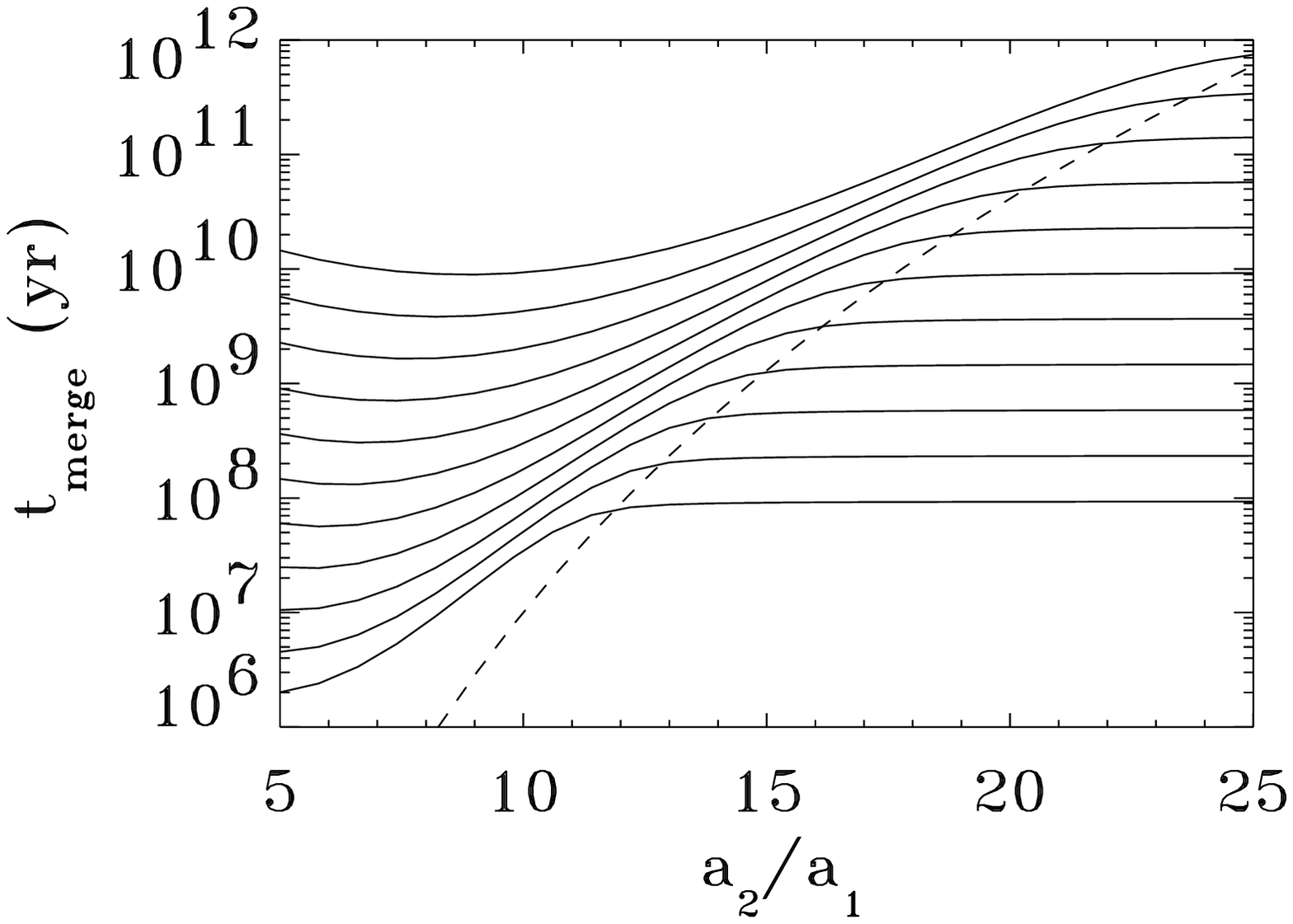}
\caption{Merger time of an inner binary black hole system with masses
$m_0=2\times10^6$~M$_\odot$ and $m_1=10^6$~M$_\odot$ in a hierarchical
triple with outer black hole mass $m_2=10^6$~M$_\odot$, as a function
of the initial semimajor axis ratio $a_2/a_1$ of the triple.  The
initial conditions of the triple are $e_1=0.1$, $e_2=0.1$, $g_1=0$,
$g_2=90^\circ$, and $i=80^\circ$. From
bottom to top, the different solid curves show results for different initial
semimajor axes of the inner binary, spaced at equal logarithmic intervals:
$a_1=\{1.00, 1.26, 1.58, 2.00, 2.51, 3.16, 3.98, 5.01, 6.31, 7.94, 10.0\}
\times10^{-3}$~pc.  The dashed curve separates the region on the left where
the Kozai resonance exists (at least initially) from that on the right where
general relativistic precession destroys the eccentricity oscillations.}
\end{figure}

Perhaps the most important initial condition parameters affecting the inner
binary merger time are the semimajor axes $a_1$ and $a_2$ and the mutual
inclination of the inner and outer orbits.  Figure 8 shows the dependence of
the merger time on the semimajor axis ratio $a_2/a_1$ for triples with
fixed initial inclination $i=80^\circ$ (the same as in all previous figures)
and various values of the initial inner semimajor axis $a_1$.  For large initial
semimajor axis ratios $a_2/a_1$ (to the right of the dashed line in figure 8),
general relativistic precession destroys
the Kozai resonance and the eccentricity of the inner binary is largely
unaffected by the outer black hole.  The merger time in this case is then
the same as that of an isolated binary, and is given by equation
(\ref{tmerge}).  On the other hand, if the outer black hole comes sufficiently
close (to the left of the dashed line in figure 8), then the Kozai resonance
exists and substantial reduction in the merger time occurs as a result of
eccentricity oscillations.  The equation for the dashed line itself which
separates these two regimes comes from using the binary merger time in equation
(\ref{tmerge}) as a proxy for the inner semimajor axis $a_1$, and employing
equation (\ref{a2a1}):
\begin{eqnarray}
t_{\rm merge}&=&1.2\times10^6{\rm yr}\left({a_2/a_1\over10}\right)^{12}
\left({2m_2\over m_0+m_1}\right)^{-4}
\left({m_0\over10^6{\rm M}_\odot}\right)^{-1}
\left({m_1\over10^6{\rm M}_\odot}\right)^{-1}\nonumber\\
& &\left({m_0+m_1\over2\times10^6{\rm M}_\odot}\right)^3
{(1-e_2^2)^6\over(1-e_1^2)^{5/2}}f(e_1).
\label{eqdash}
\end{eqnarray}

\begin{figure}[t]
\epsscale{0.6}
\plotone{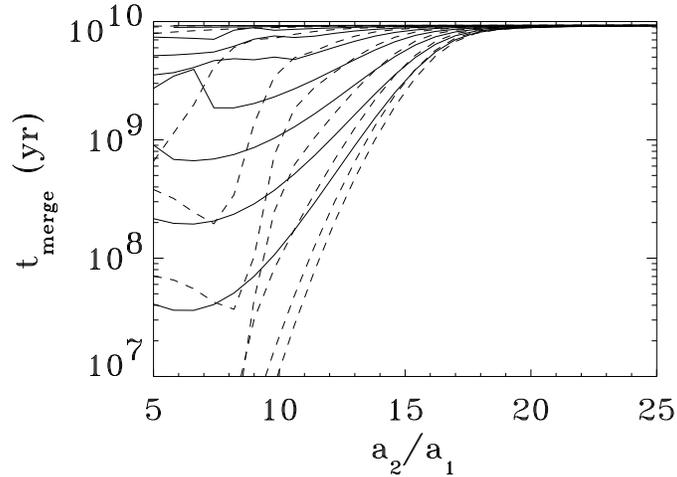}
\caption{Merger time of an inner binary black hole system with masses
$m_0=2\times10^6$~M$_\odot$ and $m_1=10^6$~M$_\odot$ in a hierarchical
triple with outer black hole mass $m_2=10^6$~M$_\odot$, as a function
of the initial semimajor axis ratio $a_2/a_1$ of the triple.  The different
curves correspond to different initial mutual inclinations of the binary: from
bottom to top, $|\cos i|$ ranges from 0.1 to 0.9 in steps of 0.1.  (The 0.8
and 0.9 curves lie almost on top of each other.)  Prograde outer orbits
($\cos i>0$) are shown by solid curves, while retrograde outer orbits
($\cos i<0$) are shown by dashed curves.
The other initial conditions of the triple are $a_1=3.16\times10^{-3}$~pc,
$e_1=0.1$, $e_2=0.1$, $g_1=0$, and $g_2=90^\circ$.  The irregular behavior
at intermediate inclinations and low values of $a_2/a_1$ is a result of
chaos.
}
\end{figure}

Figure 9 shows how the variation of merger time with initial semimajor
axis ratio depends on the initial mutual inclination
angle of the triple for initial $a_1=3.16\times10^{-3}$~pc.
As expected, significant reduction in the merger time can only occur when the 
initial inclination is high enough that the
Kozai resonance is present: $|\cos i|\la (3/5)^{1/2}\simeq0.77$.
If this criterion is satisfied, then values of $a_2/a_1$
satisfying equation (\ref{a2a1}), i.e. $a_2/a_1<17.7$ for this particular
triple, will exhibit accelerated mergers, with higher inclinations showing
the fastest merger times.  As Miller \& Hamilton (2002) point out
for the stellar mass case, retrograde triples ($\cos i<0$) generally
produce faster
merger times, at least for nearly equal masses.

\subsection{Merger Times for Other Combinations of Masses}

As we noted above in section 3, our numerical results for the nearly equal
mass triples of the previous subsection can be scaled to all triples
with the same mass ratio, $m_0:m_1:m_2=2:1:1$, provided the initial semimajor
axes are scaled by the same factor as the masses.  The merger time then scales
by exactly the same factor.  We have also investigated triples consisting
of substantially unequal masses, and our results may also be scaled to
other, similar mass ratio triples in the same manner.

\begin{figure}[t]
\epsscale{0.6}
\plotone{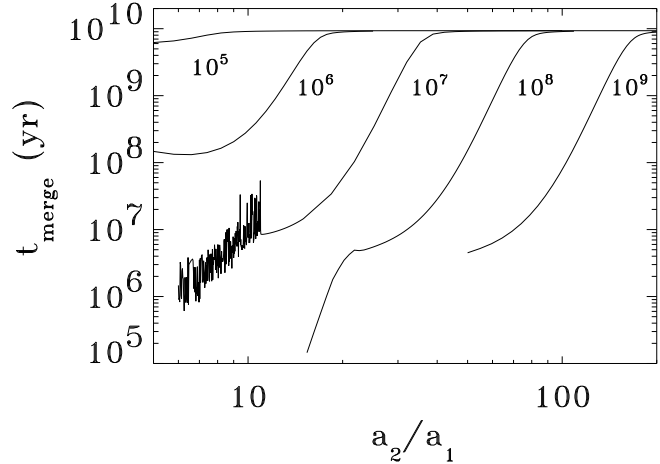}
\caption{Merger time of an inner binary black hole system with masses
$m_0=2\times10^6$~M$_\odot$ and $m_1=10^6$~M$_\odot$ in a hierarchical
triple, as a function of the initial semimajor axis ratio $a_2/a_1$ of the
triple.  The different curves correspond to different outer black hole masses,
as labeled in solar masses.  The initial
conditions of the triple are $a_1=3.16\times10^{-3}$~pc, $e_1=0.1$,
$e_2=0.1$, $g_1=0$, $g_2=90^\circ$, and $i=80^\circ$.}
\end{figure}

Figure 10 depicts the merger time for the same inner binary as shown in figure
8, but with different outer black hole masses.  All cases shown start with
an initial inner semimajor axis of $a_1=3.16\times10^{-3}$~pc, for which the
merger time would be $\simeq9.3\times10^{9}$~yr if the binary were isolated.
Because we do not account for gravitational wave losses associated with the
outer binary, all the calculations shown in the figure have $a_2/a_1$ large
enough so that the nominal gravitational wave merger time of the outer binary
from equation (\ref{tmerge}) is at least ten times longer than that of the
inner binary.

The $10^5$~M$_\odot$ curve shows that an outer black hole with a much smaller
mass than that of the inner binary does not have a
substantial effect on the merger time of that binary.  On the other hand,
a larger outer black hole mass exerts a stronger tidal perturbation
on the inner binary.  This reduces the merger time significantly compared to
the nearly equal mass case, and also does it at larger semimajor axis
ratios.  In agreement with equation (\ref{a2a1}), the value of $a_2/a_1$
required for the onset of the Kozai resonance scales with the outer black
hole mass as $m_2^{1/3}$.  Triples in which the outer black hole mass
is much larger than that of the inner binary might arise from the merger of a
smaller galaxy containing a stalled binary with a larger galaxy
containing a bigger black hole.

\begin{figure}[t]
\epsscale{0.6}
\plotone{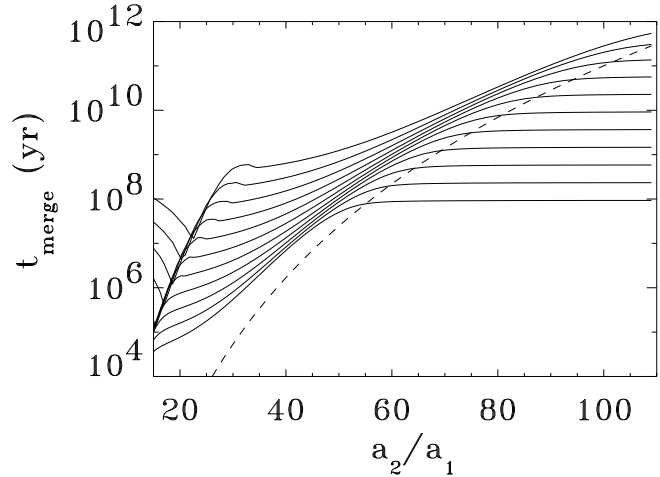}
\caption{Same as figure 8 except for an outer black hole mass of
$m_2=10^8$~M$_\odot$.}
\end{figure}

Figure 11 presents a more detailed look at the dependence of the merger time
on the semimajor axes $a_1$ and $a_2$ for the case of an outer black hole
mass of $m_2=10^8$~M$_\odot$.  With the exception of the
larger outer black hole mass, the initial triple parameters are identical to
those of figure 8.  Note the change in scales on the axes:  the larger
outer black hole mass greatly reduces the merger time at much larger values
of $a_2/a_1$.  The dashed line comes once again from equation (\ref{eqdash}),
and separates the region on the left where the Kozai resonance exists from
that on the right where relativistic periastron precession destroys the
resonance.

\begin{figure}[t]
\epsscale{0.6}
\plotone{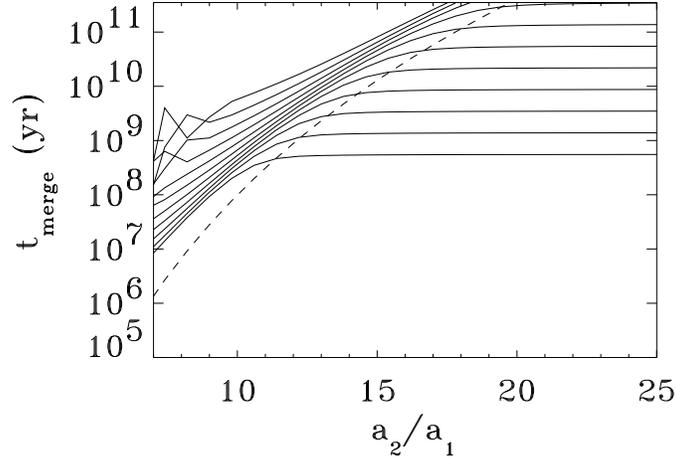}
\caption{Merger time of an inner binary black hole system with masses
$m_0=10^8$~M$_\odot$ and $m_1=10^6$~M$_\odot$ in a hierarchical
triple with outer black hole mass $m_2=10^8$~M$_\odot$, as a function
of the initial semimajor axis ratio $a_2/a_1$ of the triple.  The
initial conditions of the triple are $e_1=0.1$, $e_2=0.1$, $g_1=0$,
$g_2=90^\circ$, and $i=80^\circ$. From
bottom to top, the different solid curves show results for different initial
semimajor axes of the inner binary, spaced at equal logarithmic intervals:
$a_1=\{1.00, 1.26, 1.58, 2.00, 2.51, 3.16, 3.98, 5.01, 6.31, 7.94, 10.0\}
\times10^{-2}$~pc.  (These values of $a_1$ are ten times those of figures 8
and 11.)  The Kozai resonance exists initially only to the left of the
dashed line.}
\end{figure}

Figure 12 shows results for a substantially unequal mass inner binary consisting
of $10^8$ and $10^6$~M$_\odot$ black holes, in a triple with an outer
$10^8$~M$_\odot$ black hole.  We choose initial inner semimajor axes $a_1$ to
be ten times larger than those of figures 8 and 11 so that
$t_{\rm merge,binary}$ is roughly the same.  Very substantial reductions
in the merger time
occur once again, but for semimajor axis ratios $a_2/a_1$ that are even
smaller than in the roughly equal mass case.  Larger outer black hole masses
can expand this range of $a_2/a_1$ (see equation
[\ref{a2a1}]), but
would also accelerate the gravitational wave merger of the outer binary,
increasing the likelihood of an unstable encounter.

We have only shown results here for prograde orbits in triples with
substantially unequal masses.  In contrast to the nearly equal mass case,
retrograde orbits do not produce merger times which are very different
from their prograde counterparts.  This asymmetry between
prograde and retrograde orbits arises from terms in the equations of motion
which depend on odd powers of $\theta$, which are neglected in the usual
Kozai approximation (see Appendix A).  These terms are also very small
for the unequal mass cases we have calculated here.

\clearpage

\section{DISCUSSION AND CONCLUSIONS}

We have shown that in cases where a bound hierarchical black hole triple forms,
substantial
reductions in the gravitational wave merger time of the inner binary can
take place for reasonably large ranges of initial conditions of the triple.
For example, figure 9 shows that for a triple consisting of
nearly equal mass black holes, the merger time can be reduced by more than
a factor of ten in over fifty percent of all cases, provided the mutual
inclination
angle is randomly distributed in solid angle.  However, we have not addressed
the issue of what the distribution of initial conditions is likely to be, as
that would entail following the dynamical formation of the triple itself by
interactions with the surrounding stars and gas.  As we have shown in the
previous section, all the initial parameters of the triple affect whether
and by how much the merger of the inner binary is altered by the presence
of the outer black hole.  However, the most important are the mutual inclination
angle $i$ and the semimajor axes $a_1$ and $a_2$.  There is very little
effect on the merger time of the inner binary unless the semimajor axis ratio
$a_2/a_1$ is small enough that general relativistic precession does not destroy
the Kozai resonance (equation [\ref{a2a1}]).

An issue related to the uncertainty in initial conditions is the fact that
we have not considered the evolution of the semimajor axis of the outer black
hole due either to interactions with surrounding material or to gravitational
radiation.  The latter is not significant for the calculations shown
here, as we have always chosen $a_2/a_1$ and $e_2$ to be such that the
gravitational wave evolution time scale of the outer black hole is much longer
than the merger time scale of the inner binary.  Note that $e_2$ evolves
purely due to octupolar interactions (equation [\ref{de2dt}]), and generally
does not undergo the dramatic changes that $e_1$ exhibits.  Hence we do not
expect dramatic reductions in the {\it outer} binary merger time scale due
to interactions with the inner binary.

However, that still leaves open the issue of whether the inner binary will
merge before interactions with surrounding material cause the outer black
hole to come sufficiently close that an unstable three body encounter occurs.
It would be interesting to follow the
evolution of the triple by including the time dependence of $a_2$ starting
at large enough values that the Kozai resonance is destroyed by general
relativistic periastron precession.

In addition to the Kozai mechanism, the presence of a third black hole
may affect the merger time scale of the inner binary in other ways.  In
particular, being bound in a triple forces the inner binary to ``wander''
through space, albeit in a regular orbit rather than a stochastic trajectory.
This may enhance the binary's interactions with surrounding stars beyond
what it would have had if the outer black hole were not present, and
therefore the hardening rate might be increased.

Eccentricity oscillations in the inner binary may be induced by other
sources of tidal gravitational fields besides that of an orbiting outer
black hole, e.g. nearby matter inhomogeneities or an aspherical distribution
of surrounding stars and gas.  In order to beat general relativistic
periastron precession, equation (\ref{a2cubed}) implies that the quadrupole
moment $\rho_{2m}$ of these exterior matter distributions must satisfy
\begin{equation}
\rho_{2m}\sim{m_2\over4\pi a_2^3}>2\times10^6{\rm M}_\odot~{\rm pc}^{-3}
\left({m_0+m_1\over
2\times10^6{\rm M}_\odot}\right)^2\left({a_1\over10^{-2}{\rm pc}}\right)^{-4}
(1-e_1^2)^{-3/2}.
\end{equation}
Note that this critical quadrupole moment has a strong scaling with the binary
semimajor axis $a_1$, and it may be possible to achieve given the actual
stellar mass densities observed in galactic nuclei (e.g. Faber et al. 1997).

\acknowledgments
We thank E. Ford, S. Hughes, M. C. Miller, N. Murray, S. J. Peale,
S. Tremaine,
and Y. Wu for useful discussions, and the referee for suggestions that
greatly improved this
paper.  This work was supported in part by NASA grants NAG5~7075 and
NAG5~3646, and NSF grant AST~9970827.  We also acknowledge the hospitality
of the Canadian Institute for Theoretical Astrophysics, where part of this
work was completed.

\appendix

\section{APPENDIX: ANALYTIC ESTIMATES NEGLECTING OCTUPOLE TERMS}

In this appendix we present analytic calculations which are useful
in understanding how general relativistic precession modifies the Kozai
mechanism. We neglect the octupole terms in
the hierarchical triple evolution equations (\ref{da1dt})-(\ref{dhdt})
throughout this
section.  The only parameter of the outer binary which then changes
is $g_2$, and this evolution does not affect the inner binary.

Neglecting gravitational radiation, the evolution of the
inner binary is determined solely by equations (\ref{dg1dt}) and (\ref{de1dt}):

\begin{eqnarray}
{dg_1\over dt}&=&6C_2\left\{{1\over G_1}[4\theta^2+(5\cos2g_1-1)(1-e_1^2-
                 \theta^2)]+{\theta\over G_2}[2+e_1^2(3-5\cos2g_1)]\right\}
                 \nonumber\\
              & &+{3\over c^2a_1(1-e_1^2)}\left[{G(m_0+m_1)\over a_1}
                 \right]^{3/2},
\label{dg1dtb}
\end{eqnarray}
and
\begin{equation}
{de_1\over dt}=30C_2{e_1(1-e_1^2)\over G_1}(1-\theta^2)\sin2g_1.
\label{de1dtb}
\end{equation}

Kozai (1962) has provided an approximate analytic solution to these quadrupole
evolution equations when all general relativistic effects are neglected.
The approximation neglects the term multiplied by $\theta/G_2$ in equation
(\ref{dg1dtb}) which is smaller than the first term by
$G_1/G_2\sim(a_1/a_2)^{1/2}$ for comparable masses.
Also, equation (\ref{eqtheta}) implies that
\begin{equation}
2G_1G_2\theta+G_1^2=H^2-G_2^2,
\end{equation}
which is a constant when octupole terms and gravitational radiation are
neglected.  Again taking $G_2>>G_1$, this gives an approximate integral
of motion
\begin{equation}
\Theta\equiv(1-e_1^2)\theta^2\simeq{m_0+m_1\over Ga_1m_0^2m_1^2}
\left({H^2-G_2^2\over2G_2}\right)^2.
\label{eqTheta}
\end{equation}

For $\Theta<3/5$, Kozai's solutions exhibit two classes of dynamical
behavior in the evolution of the argument of perihelion $g_1$: libration about
$g_1=90^\circ$ or $270^\circ$, and circulation.  There then exists a critical
inclination angle for which the librating solutions degenerate to a
fixed point in the $e_1$ vs. $g_1$ phase space:
$\cos^2i_{\rm crit}=3(1-e_1^2)/5$.  For $\Theta>3/5$, the resonance
vanishes and only circulating solutions exist.  In this case the amplitude
of eccentricity oscillations is quite small, so the existence of the resonance
can be used as a necessary criterion for determining when large amplitude
eccentricity oscillations become possible.
Because $\Theta<3/5$ implies that $\cos^2i<3/5/(1-e_1^2)$, large amplitude
oscillations are only possible for high inclinations: $39^\circ\simeq
\cos^{-1}(3/5)^{1/2}<i<180^\circ-\cos^{-1}(3/5)^{1/2}\simeq141^\circ$.

General relativistic precession is itself circulation in $g_1$, so general
relativity would be expected to reduce the parameter space where librations
exist.  Equation (\ref{de1dtb})
implies that the fixed point, if it exists, still occurs at $g_1=90^\circ$ or
$270^\circ$.  If we continue to neglect the term multiplied by $\theta/G_2$ in
equation (\ref{dg1dtb}), then setting $dg_1/dt=0$ gives

\begin{equation}
\cos^2i_{\rm crit}={3\over5}(1-e_1^2)-
        {4G(m_0+m_1)^2a_2^3(1-e_2^2)^{3/2}\over5c^2m_2a_1^4(1-e_1^2)^{1/2}}.
\label{icrit}
\end{equation}

As expected, general relativistic precession decreases $\cos^2i_{\rm crit}$,
so that the fixed point can only exist if the inclination angle is pushed
higher.  This can only be achieved so long as the right hand side of equation
(\ref{icrit}) remains positive, implying that
\begin{equation}
{a_2^3\over a_1^3}<{3c^2m_2a_1(1-e_1^2)^{3/2}\over4G(m_0+m_1)^2(1-e_2^2)^{3/2}},
\label{a2cubed}
\end{equation}
which gives equation (\ref{a2a1}).

Equation (\ref{icrit}) can be rearranged to give an equation for the
eccentricity $e_{1,0}$
at the fixed point for triples with specified total angular momentum $H$,
and therefore specified Kozai constant $\Theta$ within the Kozai approximation,
\begin{equation}
1-e_{1,0}^2=\left[{5\Theta\over3}+{4G(m_0+m_1)^2a_2^3(1-e_2^2)^{3/2}
(1-e_{1,0}^2)^{1/2}\over3c^2m_2a_1^4}\right]^{1/2}.
\label{1me12}
\end{equation}
Solving this equation for $e_{1,0}$ requires solving a quartic, but written
in this way it is again obvious that general relativistic precession lowers
the value of $e_{1,0}$ for triples of given angular momentum, thereby
increasing the circulating portion of phase space at the expense of libration.

\end{document}